\documentclass[a4paper,11pt]{article}
\pdfoutput=1 

\usepackage{jheppub} 

\usepackage[T1]{fontenc} 

\usepackage{amsthm,graphicx,amsfonts,amsmath,amssymb,latexsym,cancel,slashbox}

 \theoremstyle{plain}
 \newtheorem {hypo1}{\bf\hspace{-\parindent}Hypothesis}
 \newtheorem {prop}[hypo1]{Proposition}
 \newtheorem* {plainclaim}{Claim}

 \theoremstyle{definition}
 \newtheorem {eg}[hypo1]{Example}

 \newtheorem {rmk}[hypo1]{Remark}

  \newcommand{\pf}{\begin{bpf}}

 \newcommand{\pfms}{\begin{bpfms}}
 \newcommand{\epf}{\end{bpf}\hfill$\square$\vspace{0.1cm}}
 \newcommand{\epfms}{\end{bpfms}\hfill$\square$\\ }

 \newcommand\ben{\begin{equation*}}
 \newcommand\ebn{\end{equation*}}
 \newcommand\be{\begin{equation}}
 \newcommand\eb{\end{equation}}
  \newcommand\Zb{\mathbb{Z}}
 
 \newcommand\Cb{\mathbb{C}}
 \newcommand\Pb{\mathbb{P}}

\title{\boldmath Conformal field theory of Painlev\'e VI}

\author[a,b]{O. Gamayun,}
\author[a,c]{N. Iorgov,}
\author[a,c]{O. Lisovyy}

\affiliation[a]{Bogolyubov Institute for Theoretical Physics,\\
03680, Kyiv, Ukraine}
\affiliation[b]{Physics Department, Lancaster University,\\
Lancaster, LA1 4YB, United Kingdom}
\affiliation[c]{Laboratoire de Math\'ematiques et Physique Th\'eorique CNRS/UMR 7350,
 Universit\'e de Tours,\\
 37200 Tours, France}

\emailAdd{o.gamayun@lancaster.ac.uk}
\emailAdd{iorgov@bitp.kiev.ua}
\emailAdd{lisovyi@lmpt.univ-tours.fr}

\arxivnumber{arxiv:1207.0787}

 \abstract{Generic Painlev\'e VI tau function $\tau(t)$ can be interpreted as four-point correlator of primary
 fields of arbitrary dimensions in 2D CFT with $c=1$. Using AGT combinatorial representation of conformal
 blocks and determining the corresponding structure constants, we obtain full and completely
 explicit expansion of $\tau(t)$ near the singular points. After a check of this expansion, we
 discuss examples of conformal blocks arising from Riccati, Picard, Chazy and algebraic solutions of Painlev\'e~VI.}

\begin{document} 
\maketitle
\flushbottom

 \section{Introduction}
 Starting from the early seventies, Painlev\'e equations have been playing an increasingly important role
 in mathematical physics, especially in the applications to classical and quantum integrable systems
 and random matrix theory \cite{FS}, \cite{jmms}, \cite{Lukyanov2011}, \cite{smj}, \cite{twairy}--\cite{wmtb},
 \cite{Zamo_PIII}. A considerable progress has been
 made since then in the study of various analytic, asymptotic and geometric properties of Painlev\'e transcendents.
 The interested reader is referred to  \cite{clarkson,conte,fokas} for details and further references.

 The sixth Painlev\'e  equation  (PVI)
 \begin{align}\label{pvi}
 &\,w''=\frac{1}{2}\left(\frac{1}{w}+\frac{1}{w-1}+\frac{1}{w-t}\right)\left(w'\right)^2-
 \left(\frac{1}{t}+\frac{1}{t-1}+\frac{1}{w-t}\right)w'+\\
 \nonumber
 &\,+\frac{2w(w-1)(w-t)}{t^2(t-1)^2}\left(\left(\theta_{\infty}-\text{\small $\frac12$}\right)^2-\frac{\theta_0^2t}{w^2}+
 \frac{\theta_1^2(t-1)}{(w-1)^2}-
 \frac{\left(\theta_t^2-\frac14\right)t(t-1)}{(w-t)^2}\right),
 \end{align}
 is on the top of the classification of 2nd order ODEs without movable critical points. The latter property
 means that $w(t)$ is a meromorphic function on the universal cover of $\Pb^1\backslash\{0,1,\infty\}$. Four complex
 parameters $\boldsymbol{\theta}=(\theta_0,\theta_t,\theta_1,\theta_{\infty})$ and two integration constants
 form a six-dimensional PVI parameter space~$\mathcal{M}$.

 The most natural mathematical framework for Painlev\'e equations is the theory of monodromy
 preserving deformations. For example, equation (\ref{pvi}) is
 associated to rank 2 system
 \be\label{lsys}
 \frac{d\Phi}{dz}=\left(\frac{\mathcal{A}_0}{z}+\frac{\mathcal{A}_t}{z-t}+\frac{\mathcal{A}_1}{z-1}\right)\Phi,
 \eb
 with four regular singular points $0,t,1,\infty$ on $\Pb^1$.
 Traceless $2\times 2$ matrices $\mathcal{A}_{\nu}$ ($\nu=0,t,1$) are independent of $z$ and have eigenvalues $\pm\theta_{\nu}$
 that coincide with PVI parameters. Also, $\mathcal{A}_{0}+\mathcal{A}_t+\mathcal{A}_1\substack{def \\ =\\ \;}-\mathcal{A}_{\infty}=\mathrm{diag}\left\{-\theta_{\infty},\theta_{\infty}\right\}$.

 The fundamental matrix solution
 $\Phi(z)$ is multivalued on $\Pb^1\backslash\{0,1,t,\infty\}$, as its analytic continuation along
 non-contractible closed loops produces nontrivial monodromy. Full monodromy group is generated by
 three matrices $\mathcal{M}_{0,t,1}\in G=SL(2,\Cb)$ which correspond to the loops $\gamma_{0,t,1}$ in Fig.~1
 (note that $\mathcal{M}_{\infty}\mathcal{M}_1 \mathcal{M}_t \mathcal{M}_0=\mathbf{1}$). Right
 multiplication of $\Phi$ by a constant matrix gives another solution, and therefore $\{\mathcal{M}_{\nu}\}$ are fixed
 by (\ref{lsys}) only up to overall conjugation.

 As is well-known,
 Schlesinger equations of isomonodromic deformation
 of (\ref{lsys})
 \ben
 \frac{d\mathcal{A}_0}{dt}=\frac{[\mathcal{A}_t,\mathcal{A}_0]}{t},\qquad
 \frac{d\mathcal{A}_1}{dt}=\frac{[\mathcal{A}_t,\mathcal{A}_1]}{t-1}
 \ebn
 are equivalent to (\ref{pvi}). Parameter space $\mathcal{M}$ may be identified
 with $G^{3}/G$ and many questions on Painlev\'e~VI can be recast in terms of monodromy.
 This approach, characteristic for classical integrable systems in general, turns out to be quite
 successful. In particular, it represents the key element
 of the solution of PVI connection problem \cite{jimbo},
 as well as of the construction \cite{boalch,dubrovin} and classification \cite{apvi} of
 algebraic solutions.

  \begin{figure}[!h]
 \begin{center}
 \resizebox{6cm}{!}{
 \includegraphics{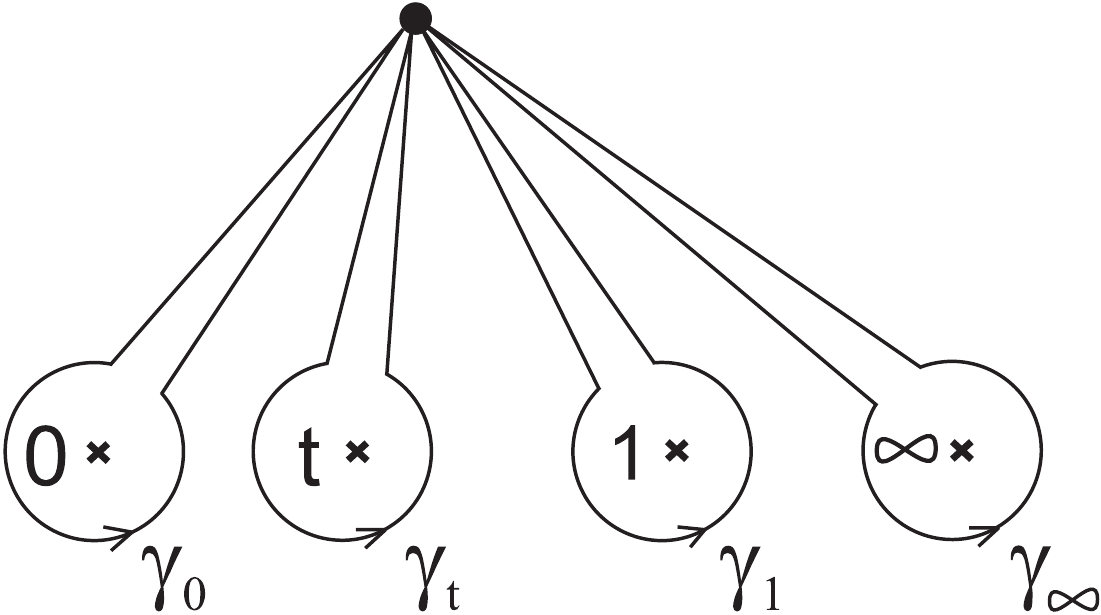}} \\
 Fig. 1
 \end{center}
 \end{figure}

 Logarithmic derivative of Painlev\'e VI tau function
 \be\label{dertau2}
 \sigma(t)=t(t-1)\frac{d}{dt}\ln\tau=(t-1)\;\mathrm{tr}\,\mathcal{A}_0\mathcal{A}_t+
 t\;\mathrm{tr}\,\mathcal{A}_t\mathcal{A}_1
 \eb
 can be expressed in terms of $t,w(t)$ and $w'(t)$ (see e.g. Eq.~(2.9) in \cite{dyson2f1}). It solves
 a nonlinear 2nd order ODE called $\sigma$-form of Painlev\'e~VI ($\sigma$PVI), which may be written as
 \begin{align}
 \label{sigmapvi}
 \Bigl(t(t-1)\sigma''\Bigr)^2=-2\;\mathrm{det}\left(\begin{array}{ccc}
 2\theta_0^2 & t\sigma'-\sigma & \sigma'+\theta_0^2+\theta_t^2+\theta_1^2-\theta_{\infty}^2 \\
  t\sigma'-\sigma & 2 \theta_t^2 & (t-1)\sigma'-\sigma \\
  \sigma'+\theta_0^2+\theta_t^2+\theta_1^2-\theta_{\infty}^2 & (t-1)\sigma'-\sigma & 2\theta_1^2
 \end{array}\right).
 \end{align}
 It is much more natural to work with $\tau(t)$ than with $w(t)$ for a number of reasons:
 \begin{itemize}
 \item
 First, it is the tau function which typically shows up in the applications of Painlev\'e equations,
 representing gap probabilities in random matrix theory, correlation functions of Ising model
 and sine-Gordon field theory at the free-fermion point etc.
 \item
 The notion of tau function extends to the general isomonodromic setting \cite{jmu} and its mathematical
 meaning is rather clear: it gives the determinant of a Cauchy-Riemann operator whose
 domain consists of multivalued functions with appropriate monodromy~\cite{palmer}.
 \item
 Thirdly and (arguably) most importantly, the tau
 function has an intimate connection with quantum field theory.
 \end{itemize}

 The last point was discovered by Sato, Miwa and Jimbo in the first two papers of the series
 \cite{smj}. There it was shown that the Riemann-Hilbert problem for rank $r$ linear systems with
 an arbitrary number of regular singularities on $\Pb^1$ admits a formal solution
 in terms of correlation functions in the theory describing $r$ free massless chiral
 fermion copies. Besides fermions,
 the correlators involve local fields of another type (below we use the term ``monodromy fields'' instead
 of SMJ's ``holonomic'') which can be seen in the operator formalism as Bogoliubov transformations
 of the fermion algebra ensuring the required monodromy properties.
  General isomonodromic tau function
 was originally \textit{defined}  as the correlator of monodromy fields. In particular, for Painlev\'e~VI
 it is given by a four-point correlator
 \be\label{taupvic}
 \tau(t)=\langle\mathcal{O}_{\mathcal{M}_0}(0)\mathcal{O}_{\mathcal{M}_t}(t)
 \mathcal{O}_{\mathcal{M}_1}(1)\mathcal{O}_{\mathcal{M}_{\infty}}(\infty)\rangle.
 \eb

 We would like to put isomonodromic deformations into the context
 of subsequent developments in conformal field theory  \cite{BPZ,DF,ZZ_book}. To our knowledge,
 no such attempt has been made so far except for a short general discussion in \cite{Moore}. The present
 paper mainly deals with Painlev\'e~VI which represents the simplest nontrivial example of isomonodromy equations.

 It will be argued below that the relevant chiral
 CFT has central charge $c=1$ and monodromy
 fields in (\ref{taupvic}) are Virasoro primaries with conformal dimensions $\Delta_{\nu}=\frac{1}{2}\,\mathrm{tr}\,\mathcal{A}_{\nu}^2=\theta^{2}_{\nu}$,
 where $\nu=0,1,t,\infty$. Therefore the structure of the expansion of $\tau(t)$ near, say, $t=0$, is strongly constrained
 by conformal invariance. In fact, the chiral correlator of four primary fields for any $c$ is given by
 \be\label{genope}
 \langle\phi_0(0)\phi_t(t)\phi_1(1)\phi_{\infty}(\infty)\rangle=\sum_p C_{0t}^pC^{1}_{p\infty}t^{\Delta_p-\Delta_0-\Delta_t}
 \mathcal{F}({\Delta},\Delta_p,c;t),
 \eb
 where the sum runs over conformal families appearing in the OPE of $\phi_0$ and $\phi_t$,
 $\Delta_p$ denotes the dimension of the corresponding intermediate primary field $\phi_p$ and
 ${\Delta}=(\Delta_0,\Delta_t,\Delta_1,\Delta_{\infty})$ is the set of external dimensions. Conformal block $\mathcal{F}(\Delta,\Delta_p,c;t)=\sum\limits_{k=0}^{\infty}\mathcal{F}_k(\Delta,\Delta_p,c)\,t^k$ associated to the
 channel $p$ is a power series  normalized as
  $\mathcal{F}_0(\Delta,\Delta_p,c)=1$. It is completely fixed by conformal symmetry \cite{BPZ}.
  The structure constants $C_{0t}^p$, $C_{p\infty}^{1}$
 combine conformal blocks into correlation functions of specific theories and should be
 obtained from another source. CFT correlators usually contain contributions of holomorphic
 and antiholomorphic conformal blocks subject to the requirement of invariance under the action of braid group
 on the positions of fields. The chiral correlators (\ref{taupvic}), (\ref{genope}) are not invariant under this
 action but transform in a natural way, induced by the Hurwitz action on monodromy matrices.

 Direct computation of the coefficients $\mathcal{F}_k(\Delta,\Delta_p,c)$ becomes
 quite laborious with the growth of $k$.
 However, very recently this problem was completely solved in the framework of
 AGT conjecture \cite{AGT} relating Liouville CFT and $\mathcal{N}=2$ 4D supersymmetric gauge theories.
 The latter correspondence produces conjectural combinatorial evaluations of conformal blocks, subsequently proven
 by Alba, Fateev, Litvinov and Tarnopolsky \cite{AGT_proof}. For $c=1$, which is the only case of interest for PVI,
 another derivation was given by Mironov, Morozov and Shakirov in \cite{MMSh1}. In particular, AGT representation expresses the contribution
 of fixed level of descendants of $\phi_p$ to 4-point conformal block $\mathcal{F}(\Delta,\Delta_p,c;t)$
 in terms of sums of simple explicit functions of
 $\Delta$, $\Delta_p$ and $c$ over bipartitions with a fixed number of boxes.

 Hence, to obtain full expansion of PVI tau function it suffices to determine
 the dimension spectrum of primaries present in the OPEs of monodromy
 fields and the associated structure constants.
 To formulate the final result, we introduce monodromy exponents $\boldsymbol{\sigma}=(\sigma_{0t},\sigma_{1t},\sigma_{01})$~by
 \ben
 p_{\mu\nu}=\mathrm{tr}\,\mathcal{M}_{\mu}\mathcal{M}_{\nu}=2\cos2\pi\sigma_{\mu\nu},\qquad \mu,\nu=0,t,1.
 \ebn
 Together with $\boldsymbol{\theta}$, these parameters define seven invariant functions on the space
 $\mathcal{M}$
 of monodromy data subject to a relation \cite{jimbo}
 \begin{align}\label{jfr}
 p_{0t}^2+p_{1t}^2+p_{01}^2+p_{0t}p_{1t}p_{01}+p_0^2+p_t^2+p_1^2+p_{\infty}^2+p_0p_tp_1p_{\infty}=\\
 \nonumber =(p_0p_t+p_1 p_{\infty})p_{0t}+(p_1p_t+p_0p_{\infty})p_{1t}+
 (p_0p_1+p_tp_{\infty})p_{01}+4,
 \end{align}
 where $p_{\nu}=\mathrm{tr}\,\mathcal{M}_{\nu}=2\cos2\pi\theta_{\nu}$ ($\nu=0,t,1,\infty$). This of course agrees
 with the dimension of $\mathcal{M}$, and allows to interpret the triple $\boldsymbol{\sigma}$ as
 a pair of PVI integration constants. Below we assume that $\boldsymbol{\theta}$, $\boldsymbol{\sigma}$ are generic complex numbers verifying
 Jimbo-Fricke relation (\ref{jfr}).

 Let $\mathbb{Y}$ be the set of all partitions identified with Young diagrams.
 Given $\lambda\in\mathbb{Y}$, we write $\lambda_i$ and $\lambda'_j$ for  the number of boxes in the $i$th row and the $j$th column of $\lambda$, and denote by $|\lambda|$ the total number of boxes in $\lambda$.
 The quantity $h_{\lambda}(i,j)=\lambda'_j-i+\lambda_i-j+1$ is called the hook length of the box $(i,j)\in \lambda$.

 Our main statement is the following
 \begin{plainclaim}
 Complete expansion of Painlev\'e VI tau function near $t=0$ can be written as
 \begin{align}
 \label{tauexp}
 \tau(t)=\mathrm{const}\cdot\sum_{n\in\Zb}C_n\left(\boldsymbol{\theta},\boldsymbol{\sigma}\right)
 t^{\left(\sigma_{0t}+n\right)^2-\theta_0^2-\theta_t^2}\mathcal{B}(\boldsymbol{\theta},\sigma_{0t}+n;t).
 \end{align}
 The function $\mathcal{B}(\boldsymbol{\theta},\sigma;t)$ is a power series in $t$ which coincides with the general
 $c=1$ conformal block $\mathcal{F}\left(\theta_0^2,\theta_t^2,
 \theta_1^2,\theta_{\infty}^2,\sigma^2,1;t\right)$
 and is explicitly given by
 \begin{align}
 \mathcal{B}\left(\boldsymbol{\theta},\sigma;t\right)=(1-t)^{2\theta_t\theta_1}\sum_{\lambda,\mu\in\mathbb{Y}}
 \mathcal{B}_{\lambda,\mu}\left(\boldsymbol{\theta},\sigma\right)   t^{|\lambda|+|\mu|},
 \end{align}
 \begin{align}\label{blm}
 \mathcal{B}_{\lambda,\mu}\left(\boldsymbol{\theta},\sigma\right)=&\,
 \prod_{(i,j)\in\lambda}
 \frac{\left(\left(\theta_t+\sigma+i-j\right)^2-\theta_0^2\right)
 \left(\left(\theta_1+\sigma+i-j\right)^2-\theta_{\infty}^2\right)}{
 h_{\lambda}^2(i,j)\left(\lambda'_j-i+\mu_i-j+1+2\sigma\right)^2}\,\times\\
 \nonumber\times &\, \prod_{(i,j)\in\mu}
 \frac{\left(\left(\theta_t-\sigma+i-j\right)^2-\theta_0^2\right)
 \left(\left(\theta_1-\sigma+i-j\right)^2-\theta_{\infty}^2\right)}{
 h_{\mu}^2(i,j)\left(\mu'_j-i+\lambda_i-j+1-2\sigma\right)^2}\,.
 \end{align}
 The structure constants $\{C_n\left(\boldsymbol{\theta},\boldsymbol{\sigma}\right)\}_{n\in\Zb}$
 can be written in terms of Barnes G-function,
 \be\label{cn}
 C_n\left(\boldsymbol{\theta},\boldsymbol{\sigma}\right)=s^n\frac{
 \prod_{\epsilon,\epsilon'=\pm}G\bigl(1+\theta_t+\epsilon\theta_0+\epsilon'(\sigma_{0t}+n)\bigr)\,
 G\bigl(1+\theta_1+\epsilon\theta_{\infty}+\epsilon'(\sigma_{0t}+n)\bigr)}{
 G\bigl(1+2(\sigma_{0t}+n)\bigr)G\bigl(1-2(\sigma_{0t}+n)\bigr)},
 \eb
 with $s$ given by 
 \begin{align}\label{spm1}
 s^{\pm1}&\left(\cos2\pi(\theta_t\mp\sigma_{0t})-\cos2\pi\theta_0\right)
 \left(\cos2\pi(\theta_1\mp\sigma_{0t})-\cos2\pi\theta_{\infty}\right)=\\
 \nonumber=&\left(\cos2\pi\theta_t\cos2\pi\theta_1+\cos2\pi\theta_0\cos2\pi\theta_{\infty}\pm i \sin2\pi\sigma_{0t}\cos2\pi\sigma_{01}\right)-\\
 \nonumber-&\left(\cos2\pi\theta_0\cos2\pi\theta_1+\cos2\pi\theta_t\cos2\pi\theta_{\infty}\mp i \sin2\pi\sigma_{0t}\cos2\pi\sigma_{1t}\right)e^{\pm 2\pi i\sigma_{0t}}.
 \end{align}
 Analogous expansions of $\tau(t)$ at $t=1,\infty$ are obtained from (\ref{tauexp})--(\ref{spm1})
 by applying a suitable transformation of parameters.
 \end{plainclaim}

 Painlev\'e transcendents are generally believed to be rather complicated special functions.
 In our opinion, this reputation is somewhat undeserved. Painlev\'e~VI, for instance, enjoys most of the basic
 properties of the Gauss hypergeometric equation. In particular, it has many elementary solutions (see \cite{apvi} and references therein), B\"acklund transformations \cite{okamoto}, quadratic Landen-type transformations \cite{kitaev_LMP}, and its connection problem is solved \cite{jimbo}. The present work adds to this
 list a link to representation theory of the Virasoro algebra
 and the series representation of PVI solutions, which can
 also be seen as an efficient tool for their numerical computation.

 The outline of the paper is as follows. The next section starts with a brief survey
 of the isomonodromy problem.
 After exhibiting global conformal symmetry of the tau function, in Subsection~\ref{subs2.3} we introduce
 monodromy fields and explain how various mathematical objects of the theory
 of monodromy preserving deformations can be written in terms of correlation functions
 in 2D CFT. Subsection~\ref{subs2.4} deals more specifically with Painlev\'e~VI. Here we present the arguments leading to (\ref{tauexp})--(\ref{spm1}), and discuss analytic continuation of PVI solutions and their B\"acklund transformations from
 CFT point of view. Section~\ref{sec3} is devoted to direct analytic verification of the above expansion.
 This task is further pursued in Section~\ref{sec4}, where the known special PVI solutions are examined from the
 field-theoretic perspective. We conclude with a list of open questions and directions for future work.

\section{CFT approach to isomonodromy}
 \subsection{General} It is instructive to start with the general case
 of rank $N$  linear system with $n$ regular singular points $a=\{a_1,\ldots,a_n\}$ on $\Pb^1$. Instead of (\ref{lsys})
 one has
 \be
 \label{dphiz}
 \partial_z\Phi=\mathcal{A}(z)\Phi, \qquad \mathcal{A}(z)=\sum_{\nu=1}^n\frac{\mathcal{A}_{\nu}}{z-a_{\nu}}.
 \eb
 The absence of singularity at infinity implies that $N\times N$ constant matrices $\{\mathcal{A}_{\nu}\}$ satisfy the constraint $\sum\nolimits_{\nu=1}^n\mathcal{A}_{\nu}=0$. They are assumed to be  diagonalizable so that $\mathcal{A}_{\nu}=\mathcal{G}_{\nu}\mathcal{T}_{\nu}\mathcal{G}_{\nu}^{-1}$
 with some $\mathcal{T}_{\nu}=\mathrm{diag}\left\{\lambda_{\nu,1},\ldots,\lambda_{\nu,N}\right\}$. The fundamental
 solution will be normalized by $\Phi(z_0)=\mathbf{1}_{N}$. It is useful to introduce the matrix
  \ben
  \mathcal{J}(z)=\Phi^{-1}\partial_z\Phi=\Phi^{-1}\mathcal{A}(z)\Phi.
  \ebn
  The coefficients of the Taylor series of $\Phi(z)$ around
  $z=z_0$ can be expressed in terms of $\mathcal{J}$ and its derivatives. In particular,
 \be\label{phiexpzz0}
 \Phi(z\rightarrow z_0)=\mathbf{1}_{N}+\mathcal{J}\left(z_0\right)\left(z-z_0\right)+
 \left(\mathcal{J}^2(z_0)+\partial\mathcal{J}(z_0)\right)\frac{\left(z-z_0\right)^2}{2}+\ldots
 \eb
 Near the singular points, the fundamental solution
 has the following expansions (under additional non-resonancy
 assumption $\lambda_{\nu,j}-\lambda_{\nu,k}\notin\mathbb{Z}$ for $j\neq k$):
 \be\label{sbeh}
 \Phi(z\rightarrow a_{\nu})=\mathcal{G}_{\nu}(z)\left(z-a_{\nu}\right)^{\mathcal{T}_{\nu}}\mathcal{C}_{\nu}.
 \eb
 Here $\mathcal{G}_{\nu}(z)$ is holomorphic and invertible in a neighborhood of $z=a_{\nu}$
 and satisfies $\mathcal{G}_{\nu}(a_{\nu})=\mathcal{G}_{\nu}$. The connection matrix $\mathcal{C}_{\nu}$ is independent of $z$. Counterclockwise continuation of $\Phi(z)$ around $a_{\nu}$ leads to monodromy matrix $\mathcal{M}_{\nu}=\mathcal{C}_{\nu}^{-1}e^{2\pi i \mathcal{T}_{\nu}}\mathcal{C}_{\nu}$.

 Let us now vary the positions of singularities and normalization point,
 simultaneously evolving $\mathcal{A}_{\nu}$'s in such a way that the monodromy is preserved.
 A classical result translates this requirement into a system of PDEs
 \begin{align}
 \label{dphia}
 \partial_{a_{\nu}}\Phi=&\,-\frac{z_0-z\;\,}{z_0-a_{\nu}}\,\frac{\mathcal{A}_{\nu}}{z-a_{\nu}}\,\Phi,\\
 \label{dphiz0}
 \partial_{z_0}\Phi=&\,-\mathcal{A}\left(z_0\right)\Phi.
 \end{align}
 It is important to note that the matrix $\mathcal{J}(z)$ remains invariant under isomonodromic variation of $z_0$.
 Schlesinger  deformation equations are obtained as
 compatibility conditions of (\ref{dphiz}), (\ref{dphia}) and (\ref{dphiz0}). Explicitly,
 \begin{align}
 \label{schl1}
 \partial_{a_{\mu}}\mathcal{A}_{\nu}=\frac{z_0-a_{\nu}}{z_0-a_{\mu}}\,
 \frac{\left[\mathcal{A}_{\mu},\mathcal{A}_{\nu}\right]}{a_{\mu}-a_{\nu}},\qquad \mu\neq\nu,\qquad\\
 \label{schl2}
 \partial_{a_{\nu}}\mathcal{A}_{\nu}=-\sum_{\mu\neq\nu}
 \frac{\left[\mathcal{A}_{\mu},\mathcal{A}_{\nu}\right]}{a_{\mu}-a_{\nu}},\qquad
 \partial_{z_{0}}\mathcal{A}_{\nu}=-\sum_{\mu\neq\nu}
 \frac{\left[\mathcal{A}_{\mu},\mathcal{A}_{\nu}\right]}{z_{0}-a_{\mu}}.
 \end{align}
 Lax form of  the Schlesinger system (\ref{schl1})--(\ref{schl2}) implies that the eigenvalues of
 $\mathcal{A}_{\nu}$'s
 are  conserved under deformation. This is of course expected due to the obvious relation
 between the spectra of  $\mathcal{A}_{\nu}$'s and monodromy matrices.

 Isomonodromic tau function $\tau(a)$ is defined by
 \be\label{taudef}
 d\ln\tau=\sum_{\mu<\nu}\mathrm{tr}\,\mathcal{A}_{\mu}\mathcal{A}_{\nu} \;d\ln\left(a_{\mu}-a_{\nu}\right).
 \eb
 It is a nontrivial  consequence of the deformation equations that the 1-form in the r.h.s. is closed.
 To show that it does not depend on $z_0$,
 one can rewrite (\ref{taudef}) as
 \be\label{dertau}
 \partial_{a_{\mu}}\ln\tau=
 \sum_{\nu\neq\mu}\frac{\mathrm{tr}\,\mathcal{A}_{\mu}\mathcal{A}_{\nu} }{a_{\mu}-a_{\nu}}=
 \frac12\, \mathrm{res}_{\,z=a_{\mu}}\,\mathrm{tr}\,\mathcal{J}^2(z).
 \eb
 Here the first equality follows from (\ref{taudef}) and the second one
 from the fact that $\mathcal{J}(z)$ is conjugate to $\mathcal{A}(z)$.

 Finally, let us decompose all $\mathcal{A}_{\nu}$'s into the sum of scalar and traceless
 part as
 $ \displaystyle \mathcal{A}_{\nu}=\frac{\mathrm{tr}\,\mathcal{A}_{\nu}}{N}\,\mathbf{1}_N+\hat{\mathcal{A}}_{\nu}$.
 It can then be easily checked that
 \begin{align}
 \label{traceless1}
 \Phi_{\mathcal{A}}(z)=&\,\prod_{\nu}\Bigl(\;\frac{z-a_{\nu}}{z_0-a_{\nu}}\Bigr)^{\frac{\mathrm{tr}\,\mathcal{A}_{\nu}}{N}}
 \Phi_{\hat{\mathcal{A}}}(z),\\
  \label{traceless2}
 \mathcal{J}_{\mathcal{A}}(z)=&\,\frac1N \sum_{\nu}\frac{\mathrm{tr}\,\mathcal{A}_{\nu}}{z-a_{\nu}}\;\mathbf{1}_N\,+
 \mathcal{J}_{\hat{\mathcal{A}}}(z),\\
  \label{traceless3}
 \tau_{\mathcal{A}}(a)=&\,\prod_{\mu<\nu}\left(a_{\mu}-a_{\nu}\right)^{
 \frac{\mathrm{tr}\,\mathcal{A}_{\mu}\mathrm{tr}\,\mathcal{A}_{\nu}}{N}}\tau_{\hat{\mathcal{A}}}(a).
 \end{align}
 This allows to assume without any loss of generality that $\mathcal{A}_{\nu}$'s are traceless,
 but we deliberately postpone the imposition of this condition.

 \subsection{Global conformal symmetry} Fractional linear maps  $\displaystyle
 f(z)= \frac{\alpha z+\beta}{\gamma z+\delta}$ form the automorphism group of the Riemann sphere.
 It is clear that under the action of these transformations on $z$, $z_0$ and $a$ the quantities
 $\Phi(z)$ and $\{\mathcal{A}_{\nu}\}$ transform as functions and $\mathcal{J}(z)$ as a vector field.
 Our task in this subsection is to understand the effect of global conformal mappings on the tau function.

 Let us first compute $\tau(a)$ explicitly in the  case $n=3$. Since $\mathcal{A}_1+\mathcal{A}_2+
 \mathcal{A}_3=0$, the coefficients $\mathrm{tr}\,\mathcal{A}_{\mu}\mathcal{A}_{\nu}$ in (\ref{taudef})
 are conserved quantities. Hence, integrating (\ref{taudef}), we find
 \ben
 \tau(a_1,a_2,a_3)=\mathrm{const}\cdot\left(a_1-a_2\right)^{\Delta_3-\Delta_2-\Delta_1}
 \left(a_1-a_3\right)^{\Delta_2-\Delta_1-\Delta_3}
 \left(a_2-a_3\right)^{\Delta_1-\Delta_2-\Delta_3},
 \ebn
 where $\Delta_{\nu}=\frac12\,\mathrm{tr}\,\mathcal{A}_{\nu}^2$ with $\nu=1,2,3$. One recognizes here
 the general expression for the three-point correlation function of (quasi)primary fields of dimensions
 $\Delta_{1,2,3}$ in the two-dimensional conformal field theory.

 Given the above example, it is natural
 to assume that for general $n$ the tau function transforms as the $n$-point function of primaries
 with appropriate dimensions:
 \be\label{tautr}
 \tau\left(f(a)\right)=\prod_{\nu=1}^n\left[f'\left(a_{\nu}\right)\right]^{-\Delta_{\nu}}\,\tau(a).
 \eb
 To prove the last formula, it is sufficient to consider  infinitesimal transformations generated by
 the vector field $Z=\left(A+Bz+Cz^2\right)\partial_{z}$, which amounts to checking three differential
 constraints:
 \begin{align*}
 &\sum_{\nu}\partial_{a_{\nu}}\ln\tau=0,\\
 &\sum_{\nu}\left(a_{\nu}\partial_{a_{\nu}}\ln\tau+\Delta_{\nu}\right)=0,\\
 &\sum_{\nu}\left(a_{\nu}^2\partial_{a_{\nu}}\ln\tau+2\Delta_{\nu}a_{\nu}\right)=0.
 \end{align*}
 These relations can indeed be straightforwardly demonstrated using the first equality in (\ref{dertau}) and
 the condition $\sum_{\nu}\mathcal{A}_{\nu}=0$.

 \subsection{Field content}\label{subs2.3} The fundamental solution $\Phi$ is completely fixed by
 its monodromy, normalization and singular behaviour (\ref{sbeh}).
 The last point is particularly important:
 indeed, adding an integer to any diagonal element of $\mathcal{T}_{\nu}$ modifies
 the asymptotics of $\Phi$ as $z\rightarrow a_{\nu}$ without changing monodromy matrices.
 Therefore, in what follows, the notion of monodromy will include not only the set
 of $\mathcal{M}_{\nu}$'s
 but also the choice of their logarithm branches $\mathcal{L}_{\nu}=\mathcal{C}_{\nu}^{-1}\mathcal{T}_{\nu}
 \mathcal{C}_{\nu}$.

 Let us now try to construct a formal QFT solution of the isomonodromic deformation problem,
 extending the ideas of \cite{Moore}, \cite{smj}$_{I-II}$. The starting point will be the following
 ansatz for $\Phi$:
 \be\label{phians}
 \Phi_{jk}(z)=(z-z_0)^{2\Delta}\frac{\langle\mathcal{O}_{\mathcal{L}_{1}}(a_1)\ldots
 \mathcal{O}_{\mathcal{L}_{n}}(a_n)\bar{\varphi}_j(z_0)\varphi_k(z)\rangle}{\langle\mathcal{O}_{\mathcal{L}_{1}}(a_1)\ldots
 \mathcal{O}_{\mathcal{L}_{n}}(a_n)\rangle},\qquad j,k=1,\ldots,N.
 \eb
 Here it is assumed that
 $\{\mathcal{O}_{\mathcal{L}_{\nu}}\}$, $\{\bar{\varphi}_j\}$, $\{\varphi_k\}$ are primary fields
 in a 2D CFT characterized by some central charge $c$. Further, we want the OPEs of $\bar{\varphi}$'s with
 $\varphi$'s to contain the identity operator.
 This forces them to have equal dimensions, to be denoted
 by $\Delta$.  Normalization of these fields is fixed by the normalization of $\Phi$; the leading
 OPE term should be equal to
 \be\label{phiphiope}
 \bar{\varphi}_j(z_0)\varphi_k(z)\sim (z-z_0)^{-2\Delta}\,\delta_{jk}.
 \eb
 Since $\Phi(z)$ may be represented by an entire series near $z=z_0$,
 the dimensions of all other primaries appearing
 in this OPE should be given by strictly positive integers.
  Monodromy fields $\{\mathcal{O}_{\mathcal{L}_{\nu}}\}$ are defined by the condition that their complete OPEs
 with $\{\varphi_k\}$ have the form
 \ben
 \mathcal{O}_{\mathcal{L}_{\nu}}(a_{\nu})\varphi_k(z)=\sum_{j=1}^n
 \left(\left(z-a_{\nu}\right)^{\mathcal{L}_{\nu}}\right)_{jk}
 \sum_{\ell=0}^{\infty}
 \mathcal{O}_{\mathcal{L}_{\nu},j,\ell}(a_{\nu})\left(z-a_{\nu}\right)^{\ell},
 \ebn
 where $\{\mathcal{O}_{\mathcal{L}_{\nu},j,\ell}\}$ are some local fields. In particular, the raw vector
 $\left(\varphi_1\ldots \varphi_n\right)$ should be multiplied by $\mathcal{M}_{\nu}$ when continued around
 $\mathcal{O}_{\mathcal{L}_{\nu}}$. If one succeeds in finding
 a set of fields with all mentioned properties, the correlator ratio (\ref{phians}) will automatically
 give the solution of the linear system (\ref{dphiz}).

 The definition (\ref{taudef}) of the tau function arises very naturally in
 the CFT framework. To illustrate this, let us compute two more orders in the OPE (\ref{phiphiope}). The identity
 field has no level~1 descendants, therefore the leading correction is given by a new primary field
 $J_{jk}$ of dimension~1. The next-to-leading order
 correction comes from three sources:
 i) nonvanishing level 2 descendant of the identity operator
 given by the energy-momentum tensor $T$,
 ii) level 1 current descendant $\partial J_{jk}$
 and iii) new primaries of dimension~2 which can be combined into a single field $S_{jk}$.
 Thus
 \begin{align}
 \label{phiphiope2}
 \bar{\varphi}_j(z_0)\varphi_k(z)= (z-z_0)^{-2\Delta}\,\Bigl[\,\delta_{jk}+J_{jk}(z_0)\left(z-z_0\right)+\quad\\ \nonumber+\Bigl(\frac{4\Delta}{c}\,
 T(z_0)\delta_{jk}+(\partial J_{jk})(z_0)+S_{jk}(z_0)\Bigr)\frac{(z-z_0)^2}{2}+O\left((z-z_0)^3\right)\Bigr]\,.
 \end{align}
 We will make a further assumption of tracelessness of $S$, which is essentially
 motivated by the examples considered below.
 Now, substituting (\ref{phiphiope2}) into (\ref{phians}) and matching the result with (\ref{phiexpzz0}), one
 finds that
 \begin{align}
 \label{jans}
 \mathcal{J}(z)=&\,\frac{\langle\mathcal{O}_{\mathcal{L}_{1}}(a_1)\ldots
 \mathcal{O}_{\mathcal{L}_{n}}(a_n)J(z)\rangle}{\langle\mathcal{O}_{\mathcal{L}_{1}}(a_1)\ldots
 \mathcal{O}_{\mathcal{L}_{n}}(a_n)\rangle},\\
 \label{trj2ans}
 \mathrm{tr}\,\mathcal{J}^2(z)=&\,
 \frac{\langle\mathcal{O}_{\mathcal{L}_{1}}(a_1)\ldots
 \mathcal{O}_{\mathcal{L}_{n}}(a_n)T(z)\rangle}{\langle\mathcal{O}_{\mathcal{L}_{1}}(a_1)\ldots
 \mathcal{O}_{\mathcal{L}_{n}}(a_n)\rangle}\frac{4N\Delta}{c}.
 \end{align}
 Standard CFT arguments allow to rewrite the r.h.s. of the last formula  as
 \ben
 \frac{\langle\mathcal{O}_{\mathcal{L}_{1}}(a_1)\ldots
 \mathcal{O}_{\mathcal{L}_{n}}(a_n)T(z)\rangle}{\langle\mathcal{O}_{\mathcal{L}_{1}}(a_1)\ldots
 \mathcal{O}_{\mathcal{L}_{n}}(a_n)\rangle}=\sum_{\nu=1}^n\left\{\frac{\tilde{\Delta}_{\nu}}{(z-a_{\nu})^2}+
 \frac{1}{z-a_{\nu}}\,\partial_{a_{\nu}}\ln\left\langle\mathcal{O}_{\mathcal{L}_{1}}(a_1)\ldots
 \mathcal{O}_{\mathcal{L}_{n}}(a_n)\right\rangle\right\},
 \ebn
 where $\tilde{\Delta}_{\nu}$ denotes conformal dimension of
 $\mathcal{O}_{\mathcal{L}_{\nu}}$. Comparison of
 (\ref{trj2ans}) with (\ref{dertau}) then shows that the tau function can be identified
 with a power of the correlator of monodromy fields,
 \be\label{cor}
 \tau(a)=\left\langle\mathcal{O}_{\mathcal{L}_{1}}(a_1)\ldots
 \mathcal{O}_{\mathcal{L}_{n}}(a_n)\right\rangle^{\frac{2N\Delta}{c}}.
 \eb
 In what follows, we will be exclusively interested in the case when
 \be\label{cconstraint}
 c=2N\Delta.
 \eb
 Such a condition implies, in particular, that the dimensions $\tilde{\Delta}_{\nu}$ of monodromy fields
 coincide with the quantities $\Delta_{\nu}=\frac12\,\mathrm{tr}\,\mathcal{A}_{\nu}^2$ from the previous subsection.

 One possible realization of the above conditions is provided by the theory of
 $N$ free complex fermions. Its central charge $c=N$ agrees with the
 conformal dimension $\Delta=\frac12$ of fermionic fields $\{\bar{\psi}_{j}\}$, $\{\psi_k\}$ which play the role of
 $\bar{\varphi}$'s and $\varphi$'s. The currents are by definition given by
 $J_{jk}=(\bar{\psi}_j\,\psi_k)$, while the energy-momentum tensor $T$
 and the fields $\{S_{jk}\}$ may be expressed as
  \begin{align*}
  &T=\frac12\sum_{k}\left[(\bar{\psi}_k\,\partial\psi_k)-(\partial\bar{\psi}_k\,\psi_k)\right],\\
  &S_{jk}= (\bar{\psi}_{j}\,\partial\psi_k)-(\partial\bar{\psi}_{j}\,\psi_k)-\frac{2}{N}\,T\,\delta_{jk}.
  \end{align*}
  To represent monodromy fields, recall the usual bosononization formulas
  \begin{align*}
  &\bar{\psi}_k=\, :e^{-i\phi_k}:,\qquad \psi_k=\, :e^{i\phi_k}:,\\
  &J_{jk}=\begin{cases} :e^{i(\phi_k-\phi_j)}:,& j\neq k, \\
  i\,\partial\phi_k, & j=k, \end{cases} \\
  &T=-\frac12\sum_k\left(\partial\phi_k\,\partial\phi_k\right),
  \end{align*}
  where $\{\phi_k\}_{k=1,\ldots,N}$ are free complex bosonic fields with the propagator $\langle\phi_k(w)\phi_k(z)\rangle
  \sim\,-  \ln(z-w)$. Also note that for $\mathcal{C}\in GL(N,\Cb)$, monodromy matrices for the
  linearly transformed fermions
  \ben
  \bar{\psi}'_j=\sum_k \mathcal{C}_{jk}\bar{\psi}_k,\qquad\psi'_j=\sum_k \mathcal{C}^{-1}_{kj}\psi_k,
  \ebn
  are obtained from $\mathcal{M}_{\nu}$'s by conjugation by $\mathcal{C}$. In particular, setting $\mathcal{C}=
  \mathcal{C}_{\nu}$, one obtains fermions $\{\bar{\psi}^{(\nu)}_k\}$, $\{{\psi}^{(\nu)}_k\}$ with diagonal
  monodromy around $a_{\nu}$. Denote by $\{\phi^{(\nu)}_k\}$ bosonic fields associated to this ``diagonal'' fermionic basis, then monodromy field $\mathcal{O}_{\mathcal{L}_{\nu}}$ can be written as
  \ben
  \mathcal{O}_{\mathcal{L}_{\nu}}=\, :e^{ i\sum_{k} \lambda_{\nu,k}\phi^{(\nu)}_k} :.
  \ebn
  We thus need to deal with $n$ different bosonization schemes of the same theory,
  each of them being
  adapted for representing one of the monodromy fields.
  The corresponding $N$-tuples of bosons are related
  by complicated nonlocal transformations.

   The  formulas (\ref{traceless1})--(\ref{traceless3}) are a signature of the well-known
   decomposition of fermionic
  CFT into the direct sum
  $\hat{u}(1)\oplus \hat{su}(N)_1$ of two WZW theories.
  Fermion and monodromy fields are given by products of fields from the two summands:
  \begin{align*}
  &\bar{\psi}_k=\, :e^{-{i{\phi}_0}/{\sqrt{N}}}:\otimes\,\hat{\bar{\varphi}}_k,\qquad
  {\psi}_k=\, :e^{{i\phi_0}/{\sqrt{N}}}:\otimes\,\hat{\varphi}_k,\\
  &\qquad\qquad\mathcal{O}_{\mathcal{L}_{\nu}}=\, :e^{\frac{i\,\mathrm{tr}\,\mathcal{A}_{\nu}}{\sqrt{N}}\,\phi_0}:\otimes \,\mathcal{O}_{\hat{\mathcal{L}}_{\nu}}.
  \end{align*}
  Bosonic field $\phi_0$ in the $\hat{u}(1)$ factors is expressed in terms of fields introduced
  before as $\phi_0=\frac{1}{\sqrt{N}}\sum_{k=1}^N\phi_k$. The fields $\{\hat{\bar{\varphi}}_k\}$, $\{\hat{\varphi}_k\}$
  and $\{\mathcal{O}_{\hat{\mathcal{L}}_{\nu}}\}$ live in the $\hat{su}(N)_1$ WZW theory and can
  be formally written as ordered exponentials of integrated linear combinations of $\hat{su}(N)_1$-currents. It should
  be emphasized that they are Virasoro primaries but not necessarily WZW primaries. The fields $\{\hat{\bar{\varphi}}_k\}$ and $\{\hat{\varphi}_k\}$ have  the same dimension
  $\displaystyle\Delta=\frac{N-1}{2N}$, in accordance with the central charge $c_{\hat{su}(N)_1}=N-1$ \cite{KZ}. The dimension of $\mathcal{O}_{\hat{\mathcal{L}}_{\nu}}$ is equal
  to $\frac12\,\mathrm{tr}\,\hat{\mathcal{A}}_{\nu}^2$, where as above,
  $\displaystyle\hat{\mathcal{A}}_{\nu}=\mathcal{A}_{\nu}-
  \frac{\mathrm{tr}\,\mathcal{A}_{\nu}}{N}\,\mathbf{1}_N$ stands for the traceless part of $\mathcal{A}_{\nu}$.

  Now it becomes clear that imposing the tracelessness of $\mathcal{A}(z)$ corresponds to factoring out the
  $\hat{u}(1)$ piece from the fermionic theory.
  This innocently looking procedure is in fact crucial, as it drastically reduces
  the number of primary fields in the OPEs and thus makes the computation of correlation
  functions much more efficient as compared to fermionic realization.
  Therefore, in what follows we set $\mathrm{tr}\,\mathcal{A}(z)=0$, remove
  the hats from $\hat{\mathcal{A}}_{\nu}$'s, $\hat{\mathcal{L}}_{\nu}$'s, $\hat{\bar{\varphi}}$'s and $\hat{\varphi}$'s to lighten the notation, and interpret the isomonodromic tau function
  as a correlation function of primaries with dimensions $\Delta_{\nu}$ in a CFT with $c=N-1$.

  We close this subsection with an example of application of field-theoretic machinery in the case $N=2$.
  It is somewhat distinguished from CFT point of view, since for $c=1$ the dimension
  $\Delta=\frac14$ of $\bar{\varphi}$'s and $\varphi$'s corresponds to level~2 degenerate states,
  and the dimension~1 of $\{J_{jk}\}$ is degenerate at level~3. Hence the
  correlation functions
  \begin{align*}
  \mathcal{P}_{jk}=&\,\langle\mathcal{O}_{\mathcal{L}_{1}}(a_1)\ldots \mathcal{O}_{\mathcal{L}_{n}}(a_n)\bar{\varphi}_j(z_0)\varphi_k(z)\rangle,\\
  \mathcal{Q}_{jk}=&\,\langle\mathcal{O}_{\mathcal{L}_{1}}(a_1)\ldots \mathcal{O}_{\mathcal{L}_{n}}(a_n)J_{jk}(z)\rangle,
  \end{align*}
  have to satisfy linear PDEs of order 2 and 3, fixed by Virasoro symmetry. This results into the following
  statement (cf observations made in \cite{novikov}):
  \begin{prop} Under assumption $\mathrm{tr}\,\mathcal{A}(z)=0$, the matrices
  \begin{align*}
  \mathcal{P}=(z-z_0)^{-\frac12}\tau\Phi,\qquad
  \mathcal{Q}=\tau\Phi^{-1}\partial_z\Phi,
  \end{align*}
  satisfy the differential equations
  \begin{align*}
  \partial_{zz}\mathcal{P}=&\,\left\{\frac{1}{z-z_0}\,\partial_{z_0}+\frac{1}{4\left(z-z_0\right)^2}+
  \sum_{\nu}\left(\frac{1}{z-a_{\nu}}\,\partial_{a_{\nu}}+\frac{\Delta_{\nu}}{\left(z-a_{\nu}\right)^2}
  \right)\right\}\mathcal{P},\\
  \partial_{zzz}\mathcal{Q}=&\,\left\{
  4\sum_{\nu}\left(\frac{1}{z-a_{\nu}}\,\partial_{a_{\nu}z}+
  \frac{\Delta_{\nu}}{\left(z-a_{\nu}\right)^2}\,\partial_z\right)+
  2\sum_{\nu}\left(\frac{1}{\left(z-a_{\nu}\right)^2}\,\partial_{a_{\nu}}+
  \frac{2\Delta_{\nu}}{\left(z-a_{\nu}\right)^3}\right)\right\}\mathcal{Q}.
  \end{align*}
  \end{prop}
  \pf Straightforward but tedious calculation using the relations (\ref{dphiz}), (\ref{dphia}), (\ref{dphiz0}), (\ref{dertau}) and the identity $A^2=\frac12\,\mathrm{tr}\,A^2\;\mathbf{1}_2$ verified by any traceless $2\times 2$ matrix $A$. \epf

  \subsection{Painlev\'e VI}\label{subs2.4} Recall that global conformal symmetry allows
  to fix the positions of three singular points. Painlev\'e~VI equation
  corresponds to setting $N=2$, $n=4$ and sending these three points
  to $0$, $1$ and $\infty$. The remaining singular point, $z=t$, represents
  the cross-ratio of singularities, which is preserved by M\"obius transformations.

  For $\nu=0,t,1,\infty$, let us denote by $\pm\theta_{\nu}$ the eigenvalues
  of $\mathcal{A}_{\nu}$. Preceding arguments show that PVI tau function $\tau(t)$
  defined by (\ref{dertau2}) is nothing but the four-point correlator of monodromy fields,
  \be\label{taupvic2}
 \tau(t)=\langle\mathcal{O}_{\mathcal{L}_0}(0)\mathcal{O}_{\mathcal{L}_t}(t)
 \mathcal{O}_{\mathcal{L}_1}(1)\mathcal{O}_{\mathcal{L}_{\infty}}(\infty)\rangle,
 \eb
 and that these fields are Virasoro primaries with dimensions
 $\Delta_{\nu}=\theta_{\nu}^2$ in a $c=1$ conformal field
 theory. The field at infinity should be understood according to the usual CFT prescription
 \ben
 \langle\ldots \mathcal{O}\left(\infty\right)\rangle
 \substack{def \\ =\\ \;}\lim_{R\rightarrow\infty}R^{2\Delta_{\mathcal{O}}}\langle\ldots
 \mathcal{O}\left(R\right)\rangle.
 \ebn

 It is clear that auxiliary fields $\{\varphi_k\}$ should
 have monodromy $\mathcal{M}_t\mathcal{M}_0$ around all fields in the OPE of $\mathcal{O}_{\mathcal{L}_0}$ and
 $\mathcal{O}_{\mathcal{L}_t}$. Let $e^{\pm2\pi i \sigma_{0t}}$ denote the eigenvalues of
 $\mathcal{M}_t\mathcal{M}_0$ and
 $\mathcal{C}_{0t}$ be its diagonalizing transformation. Since $\sigma_{0t}$ is defined only
 up to an integer, it is natural to expect that the set of primaries present in the OPE
 of $\mathcal{O}_{\mathcal{L}_0}$ and
 $\mathcal{O}_{\mathcal{L}_t}$ consists of an infinite number of monodromy fields $\mathcal{O}_{\mathcal{L}^{(n)}_{0t}}$ with $ n\in\Zb$ and
 \ben
 \mathcal{L}^{(n)}_{0t}=\mathcal{C}_{0t}^{-1}\left(
 \begin{array}{cc} \sigma_{0t}+n & 0 \\ 0 & -\sigma_{0t}-n\end{array}\right)\mathcal{C}_{0t},
 \ebn
 i.e. of all possible monodromy fields associated to the monodromy matrix $\mathcal{M}_t\mathcal{M}_0$.
 Taking into account that conformal dimension of $\mathcal{O}_{\mathcal{L}^{(n)}_{0t}}$ is equal to $\left(\sigma_{0t}+n\right)^2$,
 the first part of our main statement (formulas (\ref{tauexp})--(\ref{blm})) now follows from
 the general formula (\ref{genope}) and AGT combina\-to\-ri\-al representations of conformal blocks \cite{AGT}.

 The structure constants $C_n(\boldsymbol{\theta},\boldsymbol{\sigma})$ of the expansion (\ref{tauexp})
 can be determined from the so-called Jimbo asymptotic formula \cite{jimbo}, expressing the asymptotics of
 PVI tau function as $t\rightarrow0$ in terms of monodromy. In fact we have already obtained the ``easier half'' of this formula.
 E.g. if $-\frac12<\mathrm{Re}\,\sigma_{0t}<\frac12$, then (\ref{tauexp}) implies that the leading behaviour
 of $\tau(t)$ is given by
 \ben
 \tau(t\rightarrow 0)\sim \mathrm{const}\cdot t^{\sigma_{0t}^2-\theta_0^2-\theta_t^2}.
 \ebn
 Subleading asymptotics, fixing the second PVI integration constant, can be rewritten in the form of a recursion relation
 on the coefficients $C_n(\boldsymbol{\theta},\boldsymbol{\sigma})$. Namely,
 \begin{align*}
 \frac{C_{n\pm1}}{C_n}=\frac{\Gamma^2\left(1\mp2(\sigma_{0t}+n)\right)}{\Gamma^2\left(1\pm2(\sigma_{0t}+n)\right)}
 \prod_{\epsilon=\pm}\frac{\Gamma\left(1+\epsilon\theta_0+\theta_t\pm(\sigma_{0t}+n)\right)
 \Gamma\left(1+\epsilon\theta_{\infty}+\theta_1\pm(\sigma_{0t}+n)\right)}{
 \Gamma\left(1+\epsilon\theta_0+\theta_t\mp(\sigma_{0t}+n)\right)\Gamma\left(1+\epsilon\theta_{\infty}+\theta_1\mp(\sigma_{0t}+n)\right)}
 \times\\
 \times\,\frac{\left(\theta_0^2-(\theta_t\mp(\sigma_{0t}+n))^2\right)
 \left(\theta_{\infty}^2-(\theta_1\mp(\sigma_{0t}+n))^2\right)}{4\left(\sigma_{0t}+n\right)^2\left(1\pm2(\sigma_{0t}+n)\right)^2}(-s)^{\pm1},
 \end{align*}
 where $s$ is defined by (\ref{spm1}). This relation can be easily solved in terms of Barnes functions,
 with the answer given by (\ref{cn}). It is interesting to note that, up to a common multiplier and appropriately
 symmetrized $s^n$ factors, $C_n$'s essentially coincide with the chiral parts \cite{SW} of the corresponding
 structure constants in the time-like Liouville theory \cite{HMW,Zamo_Liouville}.

 \begin{rmk} The structure constants (\ref{cn}) can not be completely factorized into the products
 of three-point functions due to the presence of the parameter $s$. This is an artifact of non-trivial braid
 group action on the correlation functions of monodromy fields.

 To illustrate what we have in mind, consider the analytic continuation of $\tau(t)$
 along a closed counterclockwise contour around the branch point $t=0$. In general, such a continuation induces an action of the 3-braid group (more precisely, of
 the modular group $\Gamma(2)$) on monodromy \cite{dubrovin}.
 In the case at hand, new monodromy matrices are given by
 \ben
 \mathcal{M}_0'=\mathcal{M}_t\mathcal{M}_0\mathcal{M}_t^{-1},\qquad
 \mathcal{M}_t'=
 \left(\mathcal{M}_t\mathcal{M}_0\right)\mathcal{M}_t\left(\mathcal{M}_t\mathcal{M}_0\right)^{-1},
 \qquad \mathcal{M}_1'=\mathcal{M}_1,
 \ebn
 so that $\sigma_{0t}'=\sigma_{0t}$ and
 \begin{align}
 \label{p01p}
 p_{01}'=&\,p_0p_1+p_tp_{\infty}-p_{01}-p_{0t}p_{1t},\\
 \label{p1tp}
 p_{1t}'=& \,p_1p_t+p_0p_{\infty}-p_{1t}-p_{0t}p_{01}'.
 \end{align}
 Therefore, the change of the branch of $\tau(t)$ is encoded in the change of the structure
 constants. On the other hand, one can perform analytic continuation directly in the expansion
 (\ref{tauexp}). Up to an irrelevant overall factor, this amounts to multiplication of $C_{n}(\boldsymbol{\theta},\boldsymbol{\sigma})$ by $e^{4\pi i n\sigma_{0t}}$. Since both results
 should coincide,  the structure constants have to satisfy the functional relation
 \ben
 C_{n}(\boldsymbol{\theta},\boldsymbol{\sigma}')=
 \kappa\cdot e^{4\pi i n\sigma_{0t}}C_{n}(\boldsymbol{\theta},\boldsymbol{\sigma}),
 \ebn
 where $\kappa$ is independent of $n$. The factor $s^n$ in (\ref{cn}) is a minimal solution of this relation,
  as for $\boldsymbol{\sigma}'$ defined by (\ref{p01p})--(\ref{p1tp}) one has $s(\boldsymbol{\theta},\boldsymbol{\sigma}')=e^{4\pi i \sigma_{0t}}s(\boldsymbol{\theta},\boldsymbol{\sigma})$.
 \end{rmk}

 \begin{rmk}
 Let us denote by $\mathrm{dim}\,\lambda$ the number of standard Young tableaux
 of shape $\lambda\in\mathbb{Y}$. It coincides with the dimension of the
 irreducible representation of symmetric group $S_{|\lambda|}$ associated to $\lambda$.
 Also write $d_{\lambda}$ for the number of diagonal boxes in $\lambda$ and introduce the
 Frobenius coordinates
 \ben
 p^{\lambda}_i=\lambda_i-i,\qquad q^{\lambda}_i=\lambda'_i-i,\qquad i=1,\ldots,d_{\lambda},
 \ebn
 which give the number of boxes to the right and above the $i$th diagonal box. It is well known/easy to show
 that
 \begin{align*}
 \displaystyle\frac{\mathrm{dim}\,\lambda}{|\lambda|!}=\frac{1}{\prod_{(i,j)\in\lambda}h_{\lambda}(i,j)}=
 \frac{1}{\prod_{i=1}^{d_{\lambda}}\Gamma(p_i^{\lambda}+1)\,\Gamma(q_i^{\lambda}+1)}
 \det\biggl[\frac{1}{p_i^{\lambda}+q_j^{\lambda}+1}\biggr]_{i,j=1,\ldots,d_{\lambda}},\\
 \prod_{(i,j)\in\lambda}(i-j+z)(i-j+z')=(zz')^{d_{\lambda}}\prod_{i=1}^{d_{\lambda}}\Gamma\left[
 \begin{array}{c}p^{\lambda}_i+1+z,q^{\lambda}_i+1-z,p^{\lambda}_i+1+z', q^{\lambda}_i+1-z'\\ 1+z,1-z,1+z',1-z'\end{array}\right].
 \end{align*}
 We now recognize in (\ref{blm}) typical pieces of $z$-measures on partitions \cite{BO}. It would be nice
 to understand this coincidence conceptually with the purpose to sum up the series for $\mathcal{B}(\boldsymbol{\theta},\sigma;t)$ and~$\tau(t)$.
 \end{rmk}

 \begin{rmk}
 Painlev\'e VI equation has a large group of hidden symmetries of affine Weyl type \cite{okamoto}.
 Almost all of them are manifest in the conformal expansion (\ref{tauexp}). For instance, the change of sign of any
 of parameters $\boldsymbol{\theta}$ has no effect on the tau function, since conformal blocks depend
 only on the dimensions $\Delta_{\nu}=\theta_{\nu}^2$ and the ratios of structure
 constants (\ref{cn}) also remain invariant.

 Conformal block symmetry $(\theta_0,\theta_t)\leftrightarrow
 (\theta_{\infty},\theta_1)$  and its counterparts for expansions at $t=1,\infty$ yield further simple transformations of $\tau(t)$.
 Another, less trivial symmetry that can be found by inspection of (\ref{blm}) shifts the values of all $\boldsymbol{\theta}$
 by $\displaystyle \delta=\frac{\theta_0+\theta_t+\theta_1+\theta_{\infty}}{2}$. Additional transformations
 come from crossing symmetry. In contrast to the previous ones, they also act on $t$ by fractional linear transformations
 exchanging $0,1$ and $\infty$.

 The action of generators of the above transformations
 on the parameters $\boldsymbol{\theta}$, conformal blocks
 and tau function expansions is recorded in the following table:\vspace{0.2cm}
 \begin{center}
 \begin{tabular}{|c||c|c|c|c|c|c|}
 \hline
 & $\theta_0$ & $\theta_t$ & $\theta_1$ & $\theta_{\infty}$ & $\mathcal{B}(\boldsymbol{\theta},\sigma;t)$ & $\tau(t)$\\
 \hline\hline
 $\boldsymbol{s}_0$ & $-\theta_0$ & $\theta_t$ & $\theta_1$ & $\theta_{\infty}$ & $\mathcal{B}(\boldsymbol{\theta},\sigma;t)$ & $\tau(t)$\\
 \hline
 $\boldsymbol{s}_t$ & $\theta_0$ & $-\theta_t$ & $\theta_1$ & $\theta_{\infty}$ & $\mathcal{B}(\boldsymbol{\theta},\sigma;t)$ & $\tau(t)$\\
 \hline
 $\boldsymbol{s}_1$ & $\theta_0$ & $\theta_t$ & $-\theta_1$ & $\theta_{\infty}$ & $\mathcal{B}(\boldsymbol{\theta},\sigma;t)$ & $\tau(t)$\\
 \hline
 $\boldsymbol{s}_{\infty}$ & $\theta_0$ & $\theta_t$ & $\theta_1$ & $-\theta_{\infty}$ & $\mathcal{B}(\boldsymbol{\theta},\sigma;t)$ & $\tau(t)$\\
 \hline
 $\boldsymbol{s}_{\delta}$ & $\theta_0-\delta$ & $\theta_t-\delta$ & $\theta_1-\delta$ & $\theta_{\infty}-\delta$ & $\displaystyle
 (1-t)^{\delta_{1t}\delta}
 \mathcal{B}(\boldsymbol{\theta},\sigma;t)$ &
 $t^{\delta_{0t}\delta}(1-t)^{\delta_{1t}\delta}\tau(t)\substack{\; \\ \; \\ \; \\ \;}$\\
 \hline\hline
 $\boldsymbol{r}_{0t}$ & $\theta_{\infty}$ & $\theta_1$ & $\theta_t$ & $\theta_0$ & $\displaystyle
 \mathcal{B}(\boldsymbol{\theta},\sigma;t)$ & $ t^{\Delta_{0t}}\tau(t)
 \substack{\; \\ \; \\ \; \\ \;}$\\
 \hline
 $\boldsymbol{r}_{1t}$ & $\theta_{t}$ & $\theta_0$ & $\theta_{\infty}$ & $\theta_1$ & $(1-t)^{\Delta_{1t}}\mathcal{B}(\boldsymbol{\theta},\sigma;t)$ &
 $ (1-t)^{\Delta_{1t}}\tau(t)
 \substack{\; \\ \; \\ \; \\ \;}$\\
 \hline
 $\boldsymbol{r}_{01}$ & $\theta_{1}$ & $\theta_{\infty}$ & $\theta_{0}$ & $\theta_t$ & $(1-t)^{\Delta_{1t}}\mathcal{B}(\boldsymbol{\theta},\sigma;t)$ &
 $ t^{\Delta_{0t}}(1-t)^{\Delta_{1t}}\tau(t)
 \substack{\; \\ \; \\ \; \\ \;}$\\
 \hline\hline
 $\boldsymbol{q}_{01}$ & $\theta_{1}$ & $\theta_t$ & $\theta_{0}$ & $\theta_{\infty}$ & $\displaystyle$ &
 $ \tau(1-t)\substack{\; \\ \; \\ \; \\ \;}$\\
 \hline
  $\boldsymbol{q}_{0\infty}$ & $\theta_{\infty}$ & $\theta_t$ & $\theta_{1}$ & $\theta_{0}$ & $\displaystyle$ &
 $ t^{-2\Delta_t}\tau\left(t^{-1}\right) \substack{\; \\ \; \\ \; \\ \;}$\\
 \hline
  $\boldsymbol{q}_{1\infty}$ & $\theta_{0}$ & $\theta_t$ & $\theta_{\infty}$ & $\theta_{1}$ & $(1-t)^{\Delta_0-\Delta_t-\Delta_{\sigma}}
  \mathcal{B}\left(\boldsymbol{\theta},\sigma;\frac{t}{t-1}\right)$ &
 $ (1-t)^{-2\Delta_t}\tau\left(\frac{t}{t-1}\right) \substack{\; \\ \; \\ \; \\ \;}$\\
 \hline
 \end{tabular}
 \end{center}\vspace{0.2cm}
 Here we have introduced the notation
 \begin{align*}
 \delta_{0t}=\theta_0+\theta_t-\theta_1-\theta_{\infty},&\qquad
 \Delta_{0t}=\Delta_0+\Delta_t-\Delta_1-\Delta_{\infty},\\
 \delta_{1t}=\theta_1+\theta_t-\theta_0-\theta_{\infty},&\qquad
 \Delta_{1t}=\Delta_1+\Delta_t-\Delta_0-\Delta_{\infty},
 \end{align*}
 and $\Delta_{\sigma}=\sigma^2$.
 Transformed tau functions in the last column are  (in some cases) defined up to constant factors which depend on
 the choice of normalization of the structure constants.

 To complete the picture, it remains to understand the QFT meaning of an elementary Schlesinger transformation,
 e.g. the one shifting $\theta_0$ and $\theta_{1}$ by $\frac12$. It may be expected that this symmetry arises from
 the fusion of auxiliary fields $\bar{\varphi}_j$ and $\varphi_k$ with monodromy fields $\mathcal{O}_{\mathcal{L}_0}$
 and $\mathcal{O}_{\mathcal{L}_1}$ in the correlator representation (\ref{phians}) of the fundamental solution $\Phi$. Indeed, $\bar{\varphi}$'s and $\varphi$'s are degenerate at level~2, and therefore
 their OPEs with monodromy field of dimension $\theta^2$
 can only contain two conformal families generated by monodromy fields with dimensions $\left(\theta\pm\frac12\right)^2$.
 \end{rmk}

 \section{Painlev\'e VI recurrence}\label{sec3}
 Painlev\'e VI tau function expansion (\ref{tauexp})--(\ref{cn}) is, in the strict mathematical sense,
 a conjecture. On the other hand, $\tau(t)$ is completely fixed by its leading asymptotics
 as $t\rightarrow0$. Once the quantity $\sigma_{0t}$ in the power-law  exponent and the amplitude
 ratio $\displaystyle
 \frac{C_{1}(\boldsymbol{\theta},\boldsymbol{\sigma})}{C_{0}(\boldsymbol{\theta},\boldsymbol{\sigma})}$
 are found from Jimbo's formula, the rest of the series can, at least in principle, be recursively
 reconstructed order by order
 from $\sigma$PVI equation and checked against CFT predictions. Below we describe technical
 details of this procedure and derive several first terms of the conformal expansion.

 The equation (\ref{sigmapvi}) gives a quadrilinear 3rd order ODE for the tau function itself. Observe, however, that
 if we differentiate (\ref{sigmapvi}) one more time with respect to $t$, the resulting equation will be
 divisible by $\sigma''(t)$ and in the end turns out to be \textit{bilinear} in $\tau(t)$. It is  also convenient to
 work with a slightly modified function $\eta(t)=t^{\Delta_0+\Delta_t}(1-t)^{\Delta_t+\Delta_1}\tau(t)$,
 which partially takes into account the behaviour of $\tau(t)$ as $t\rightarrow0,1$. It satisfies the equation
 \begin{align}
 \label{pvibilinear}
 t^2(t-1)^2\Bigl\{\Bigl[t(t-1)\eta^2\Bigr]''''-8\Bigl[t(t-1)\left(\eta'\right)^2\Bigr]''-4(2A-1)\eta\eta''
 +8(A-2)\left(\eta'\right)^2\Bigr\}&+\\
 \nonumber+\, 16t^3(t-1)^3\left(\eta''\right)^2+4t(t-1)\eta\eta''-4(A-1)t(t-1)(2t-1)\eta\eta'+4(Bt+C)\eta^2&=0,
 \end{align}
 where $A$, $B$, $C$ are given by
 \begin{align*}
 A=&\;\;\Delta_0+\Delta_t+\Delta_1+\Delta_{\infty},\\
 B=&\left(\Delta_0-\Delta_1\right)\left(\Delta_{\infty}-\Delta_t\right),\\
 C=&\left(\Delta_0-\Delta_t\right)\left(\Delta_{1}-\Delta_{\infty}\right),
 \end{align*}
 and the fourth PVI parameter is killed by differentiation.

  Now substitute into (\ref{pvibilinear}) the  ansatz
 \be\label{pans}
 \eta(t)=\sum_{n\in\Zb}C_nt^{(\sigma+n)^2}\sum_{k=0}^{\infty}\eta_k^{(n)}t^k,
 \eb
 normalized by $\eta_0^{(n)}=1$ for all $n\in\Zb$. Picking up the coefficients
 of different powers of $t$ in the result, one obtains an overdetermined system of equations
 for  $\{C_n\}$, $\{\eta_k^{(n)}\}$. Namely, for any $\ell\in\Zb$ and any $L\in\Zb_{\geq0}$
 we have
 \begin{align}\label{rr}
 \sum_{\substack{m,n\in\Zb\\m+n=\ell}}\!\!\!C_m C_n\biggl[\!\!\!\!\!\!\!\!\!&\sum_{\substack{k,k'\geq0 \\ k+k'+m^2+n^2=L+1}}\!\!\!\!\!\!\!\!\!\!\!\alpha^{kk'}_{mn}\eta^{(m)}_k\eta^{(n)}_{k'}+
 \!\!\!\!\!\!\!\!\!\sum_{\substack{k,k'\geq0 \\ k+k'+m^2+n^2=L}}\!\!\!\!\!\!\!\!\beta^{kk'}_{mn}\eta^{(m)}_k\eta^{(n)}_{k'}+\\
 \nonumber+\!\!\!\!\!\!\!\!\!\!\!\!&\sum_{\substack{k,k'\geq0 \\ k+k'+m^2+n^2=L-1}}\!\!\!\!\!\!\!\!\!\!\!\gamma^{kk'}_{mn}\eta^{(m)}_k\eta^{(n)}_{k'}+
 \!\!\!\!\!\!\!\!\!\!\!\sum_{\substack{k,k'\geq0 \\ k+k'+m^2+n^2=L-2}}\!\!\!\!\!\!\!\!\!\!\delta^{kk'}_{mn}\eta^{(m)}_k\eta^{(n)}_{k'}\;\biggr]\,=\,0,
 \end{align}
 with somewhat cumbersome but completely explicit polynomial coefficients:
  \begin{align*}
 \alpha_{mn}^{kk'}=&\,\epsilon_{mn}^{kk'}
 \left(2P-\epsilon_{mn}^{kk'}+1\right),\\
 \beta_{mn}^{kk'}=&\,3\left(\epsilon_{mn}^{kk'}\right)^2-2\left(P-1+2A\right)\epsilon_{mn}^{kk'}-
 P\left(P-2A\right)+4C,\\
 \gamma_{mn}^{kk'}=&\,-3\left(\epsilon_{mn}^{kk'}\right)^2-2\left(P-4A\right)\epsilon_{mn}^{kk'}+
 (P-1)\left(P-1-2A\right)+4B,\\
 \delta_{mn}^{kk'}=&\,\epsilon_{mn}^{kk'}
 \left(2P+\epsilon_{mn}^{kk'}-3-4A\right),\\
 \epsilon_{mn}^{kk'}=&\,\Bigl(k-k'+(m-n)(\ell+2\sigma)\Bigr)^2,\\
 P=&\;\,2\sigma^2+2\sigma \ell+L.
 \end{align*}

 Let us write down the recurrence relations (\ref{rr}) and their consequences for some $L$ and~$\ell$.
 \vspace{0.2cm}\\
 \underline{$\mathbf{L=0}$, $\mathbf{\ell=0}$}: from (\ref{rr}) it follows that
 \ben
 \left(\alpha_{0,0}^{1,0}+\alpha_{0,0}^{0,1}\right)\eta_1^{(0)}+\beta_{0,0}^{0,0}=0,
 \ebn
 which in turn gives
 \ben
 \eta_1^{(0)}=-(\Delta_t+\Delta_1)+\frac{(\Delta_{\sigma}-\Delta_0+\Delta_t)(\Delta_{\sigma}-\Delta_{\infty}+\Delta_1)}{2\Delta_{\sigma}},
 \ebn
 with $\Delta_{\sigma}=\sigma^2$ as above.
 This reproduces
  $n=0$, level~1 descendant contribution to the expansion~(\ref{tauexp}). The corresponding term
 was already found in \cite{jimbo}.\vspace{0.3cm}\\
 \underline{$\mathbf{L=1}$, $\mathbf{\ell=\pm1}$}: in this case, (\ref{rr}) implies that
 \ben
 \left(\alpha_{\pm1,0}^{1,0}+\alpha_{0,\pm1}^{0,1}\right)\eta_1^{(\pm1)}+
 \left(\alpha_{\pm1,0}^{0,1}+\alpha_{0,\pm1}^{1,0}\right)\eta_1^{(0)}+
 \left(\beta_{\pm1,0}^{0,0}+\beta_{0,\pm1}^{0,0}\right)=0,
 \ebn
 and we obtain
 \ben\displaystyle \eta_1^{(\pm1)}=-(\Delta_t+\Delta_1)+
 \frac{\left(\Delta_{\sigma\pm1}-\Delta_0+\Delta_t\right)
 \left(\Delta_{\sigma\pm1}-\Delta_{\infty}+\Delta_1\right)}{2\Delta_{\sigma\pm1}}.
 \ebn
 The latter expression corresponds to level~1 descendants with $n=\pm1$.
 It coincides with $\eta_1^{(0)}$ with $\sigma$ replaced by $\sigma\pm1$, but this is not surprising: the same should be correct
 for any $n\in\Zb$ provided the conjectured periodicity of powers in (\ref{pans}) holds true.
 \vspace{0.2cm}\\
 \underline{$\mathbf{L=1}$, $\mathbf{\ell=0}$}: here we find
 \begin{align}\label{cp1m1}
 C_0^2\left[\left(\alpha_{0,0}^{2,0}+\alpha_{0,0}^{0,2}\right)\eta_2^{(0)}+
 \alpha_{0,0}^{1,1}\left(\eta_1^{(0)}\right)^2+ \left(\beta_{0,0}^{1,0}+\beta_{0,0}^{0,1}\right)\eta_1^{(0)}+\gamma_{0,0}^{0,0}\right]
 +\\ \nonumber +C_1C_{-1}\left(\alpha_{1,-1}^{0,0}+\alpha_{-1,1}^{0,0}\right)=0,
 \end{align}
 which gives level~2 descendant contribution with $n=0$:
 \begin{align*}
 &\qquad\qquad
 \eta_2^{(0)}=-\frac{(\Delta_t+\Delta_1)(\Delta_t+\Delta_1+1)}{2}-(\Delta_t+\Delta_1)\eta_1^{(0)}+\\
 &+\frac{(\Delta_{\sigma}-\Delta_0+\Delta_t)(\Delta_{\sigma}-\Delta_0+\Delta_t+1)
 (\Delta_{\sigma}-\Delta_{\infty}+\Delta_1)(\Delta_{\sigma}-\Delta_{\infty}+\Delta_1+1)}{4\Delta_{\sigma}(1+2\Delta_{\sigma})}\,+\\
 &+\frac{\left(1+2\Delta_{\sigma}\right)\left(\Delta_0+\Delta_t+
 \frac{\Delta_{\sigma}(\Delta_{\sigma}-1)-3(\Delta_0-\Delta_t)^2}{1+2\Delta_{\sigma}}\right)
 \left(\Delta_{\infty}+\Delta_1+\frac{\Delta_{\sigma}(\Delta_{\sigma}-1)-3(\Delta_{\infty}-\Delta_1)^2}{1+2\Delta_{\sigma}}\right)
 }{2\left(1-4\Delta_{\sigma}\right)^2}
 \,.
 \end{align*}
 To obtain the last formula, one should use, in addition to (\ref{cp1m1}) and the coefficients found above, the relation
 \ben
 \frac{C_1C_{-1}}{C_0^2}=\frac{\left(
 \left(\Delta_0+\Delta_t-\Delta_{\sigma}\right)^2-4\Delta_0\Delta_t\right)
 \left(
 \left(\Delta_{\infty}+\Delta_1-\Delta_{\sigma}\right)^2-4\Delta_{\infty}\Delta_1\right)
 }{16\Delta_{\sigma}^2\left(1-4\Delta_{\sigma}\right)^2}.
 \ebn
 This piece of initial data disappears after the above differentiation but can be determined from the
 quadrilinear form of $\sigma$PVI.\vspace{0.2cm}

 It is straightforward to compute more terms in  (\ref{pans}) using computer algebra. The procedure
 works as follows. For a given $L$, one should start with maximal $|\ell|$ producing a nontrivial
 relation (\ref{rr}), and then repeatedly decrease $|\ell|$ by $1$. When all possibilities are exhausted,  increase $L$
 by $1$. The coefficients determined at the first steps of this iterative procedure are listed in the table below.
 Empty entries correspond to the relations satisfied automatically or to no relations at all.\vspace{0.2cm}

  \begin{center}
  \begin{tabular}{|c||c|c|c|c|c|c|c|c|c|c|}
  \hline
  {\footnotesize\backslashbox{$\ell$}{$\;$ \\  $L$}} & $0$ & $1$ & $2$ & $3$ & $4$ & $5$ & $6$ & $7$ & $8$ & $9$ \\
  \hline\hline
  $0$ & $\eta_1^{(0)}$ & $\eta_2^{(0)}$ & $\eta_3^{(0)}$ & $\eta_4^{(0)}$
  & $\eta_5^{(0)}$ & $\eta_6^{(0)}$ & $\eta_7^{(0)}$ & $\eta_8^{(0)}$ & $\eta_9^{(0)}$ & $\eta_{10}^{(0)}$\\
  \hline
  $\pm1$ & & $\eta_1^{(\pm1)}$ & $\eta_2^{(\pm1)}$ & $\eta_3^{(\pm1)}$
  & $\eta_4^{(\pm1)}$ & $\eta_5^{(\pm1)}$ & $\eta_6^{(\pm1)}$ & $\eta_7^{(\pm1)}$ & $\eta_8^{(\pm1)}$ & $\eta_9^{(\pm1)}$\\
   \hline
  $\pm2$ & & & & $C_{\pm2}$ & $\eta_1^{(\pm2)}$ & $\eta_2^{(\pm2)}$ & $\eta_3^{(\pm2)}$
  & $\eta_4^{(\pm2)}$ & $\eta_5^{(\pm2)}$ & $\eta_6^{(\pm2)}$\\
   \hline
  $\pm3$ & & & & & & & & & $C_{\pm3}$ & $\eta_1^{(\pm3)}$\\
  \hline
  \end{tabular}
  \end{center}\vspace{0.2cm}

  In this way, we have successfully checked the expansion (\ref{tauexp})--(\ref{cn})
  for $n=0$, $\pm1$, $\pm2$, $\pm3$ going up to
  level $10$ in descendants. To give the reader an idea of the computational complexity,
  we note that there are nearly 500 bipartitions of size 10, each of them producing
  a rational function of $\boldsymbol{\theta}$, $\sigma_{0t}$
  in the corresponding expansion coefficient.

  \section{Conformal blocks and special PVI solutions}\label{sec4}
  \subsection{Riccati solutions} These PVI solutions appear
  when the monodromy of  (\ref{lsys}) is equivalent to an upper triangular one.
  Parameters $\boldsymbol{\theta}$ can be B\"acklund transformed to satisfy $\theta_0+\theta_t+\theta_1+\theta_{\infty}=0$.
  The initial conditions are also constrained and can be chosen as $\boldsymbol{\sigma}=(\theta_0+\theta_t,\theta_1+\theta_t,
  \theta_0+\theta_1)$.

  This results into a one-parameter family of PVI transcendents $w(t)$ that may be written in terms of
  Gauss hypergeometric functions,  see e.g. Proposition~49 in \cite{apvi} for explicit formulas.
  The relevant tau function, however, is extremely simple:
  \ben
  \tau(t)=\mathrm{const}\cdot t^{2{\theta_0\theta_t}}(1-t)^{2{\theta_t\theta_1}}.
  \ebn
  We recognize in the r.h.s. of this relation the four-point correlator $\langle \mathcal{V}_{\theta_0}(0)\mathcal{V}_{\theta_t}(t)\mathcal{V}_{\theta_1}(1)\mathcal{V}_{\theta_{\infty}}(\infty)\rangle$ of
  the chiral vertex operators
  $\mathcal{V}_{\theta}(z)=\,:e^{i\sqrt{2}{\theta}\phi(z)}:$
  made of free massless bosons. The contribution of conformal blocks $\mathcal{B}(\boldsymbol{\theta},\sigma_{0t}+n;t)$
  with $n\neq0$
  to the expansion (\ref{tauexp}) is annihilated by the vanishing structure constants.

  \subsection{Chazy solutions}
  As is well-known \cite{BPZ,DF,ZZ_book} and already mentioned in Section~2, the presence of degenerate states in the Virasoro module generated by
  a primary field $\phi$ leads to linear differential equations for the correlation functions involving this field.
  Possible conformal dimensions of such $\phi$'s are determined by the zeros
  of Kac determinant. They are labeled by two integers $r,s\in\Zb_{\geq1}$ and, for $c=1$, are explicitly given
  by $\Delta_{\phi_{r,s}}=\frac{(r-s)^2}{4}$. In the simplest nontrivial case
  $r=2$, $s=1$  the four-point correlator containing $\phi_{2,1}(z)$ and three arbitrary
  primaries solves a linear 2nd order ODE that can be reduced to the hypergeometric equation.

  Thus if one of the monodromy matrices $\mathcal{M}_{0,t,1,\infty}$ is equal to $-\mathbf{1}$, there exists a PVI tau
  function given by a linear combination of two solutions of the latter equation. For example, for $\theta_{\infty}=\pm\frac12$ (i.e. $\Delta_{\infty}=\frac14$)
  the general solution of
  \begin{align*}
  \tau''+2\left[\frac{\theta_0^2-\theta_1^2+\theta_t^2+\frac14}{t}+
  \frac{\theta_1^2-\theta_0^2+\theta_t^2+\frac14}{t-1}\right]&\tau'+\\
  +\biggl[\frac{(\theta_0+\theta_1)^2+\theta_t^2
  -\frac14}{t(t-1)}+\prod_{\epsilon=\pm}\Bigl(\frac{\theta_0^2+\theta_t^2-(\epsilon\theta_1+\frac{1}{2})^2}{t}+
  \frac{\theta_t^2+\theta_1^2-(\epsilon\theta_0+\frac{1}{2})^2}{t-1}\Bigr)\biggr]&\tau=0
  \end{align*}
  satisfies $\sigma$PVI. Conformal expansion of this tau function at, say, $t=0$ is determined by
  only two channels with dimensions
  $\left(\theta_1\pm\frac12\right)^2$. More precisely, one has
  \be\label{tauchazy}
  \tau(t)=\sum_{\epsilon=\pm}C_{\epsilon}t^{(\epsilon\theta_1+\frac12)^2-\theta_0^2-\theta_t^2}
  \mathcal{B}\left(\theta_0,\theta_t,\theta_1,\text{\footnotesize$\frac12$},\epsilon\theta_1+
  \text{\footnotesize$\frac12$};t\right),
  \eb
  where $C_{\pm}$ are arbitrary constants. The relevant conformal blocks can be written as
  \ben\label{cbchazy}
  \mathcal{B}\left(\theta_0,\theta_t,\theta_1,\text{\footnotesize$\frac12$},\theta_1+
  \text{\footnotesize$\frac12$};t\right)=(1-t)^{(\theta_0+\frac12)^2-\theta_t^2-\theta_1^2}\,_2F_1\left[
  \begin{array}{c}\theta_0+\theta_t+\theta_{1}+\frac12,
  \theta_0-\theta_t+\theta_{1}+\frac12 \\ 1+2\theta_1\end{array}\Bigl|\,t\right].
  \ebn
  The tau function (\ref{tauchazy})
  gives another known class of special function solutions of Painle\-v\'e~VI,
  the so-called generalized Chazy solutions (cf Lemma~33 in \cite{mazzocco2}).

  \subsection{Algebraic solutions}
  Algebraic PVI solutions correspond to finite orbits of
  the braid/mo\-du\-lar group action on monodromy of the associated linear system \cite{dubrovin}.
  All such solutions have been classified in \cite{apvi}. It turns out that there are 45
  exceptional equivalence classes with fixed rational parameters $\boldsymbol{\theta}$, $\boldsymbol{\sigma}$
  and three families continuously depending on some of them.

  In the exceptional cases, the contributions of different conformal blocks overlap in the tau function
  expansion, which makes difficult their identification. On the other hand, this task is rather
  straightforward for continuous
  families, see examples given below. An infinite number of other explicit examples
  of $c=1$ conformal blocks can be generated by  B\"acklund transformations.

  \begin{eg} For $\boldsymbol{\theta}=\left(a,a,b,\frac12-b\right)$, $\boldsymbol{\sigma}=\left(2a,\frac14,\frac14\right)$ there is a solution
  \ben
  \tau(t)=\mathrm{const}\cdot t^{2a^2}\underbrace{(1-t)^{-a^2-b^2+\frac{1}{16}}
  \left(\frac{1+\sqrt{1-t}}{2}\right)^{-4a^2+\left(2b-\frac12\right)^2}}_{\mathcal{B}\left(a,a,b,\frac12-b,2a;t\right)},
  \ebn
  arising from the contribution of a single conformal block (Solution~II in \cite{apvi}).
  \end{eg}
  \begin{eg}
  For $\boldsymbol{\theta}=\left(2a,a,a,\frac13\right)$, $\boldsymbol{\sigma}=\left(3a,\frac16,\frac14\right)$
  there is a solution
  that can be parame\-trized as follows (Solution~III in \cite{apvi}):
  \begin{align*}
  \tau(t(s))=&\;\mathrm{const}\cdot (s-2)^{4a^2}(s-1)^{2{a^2}-\frac{7}{72}}(s+1)^{-{10a^2}+\frac{1}{8}}(s+2)^{{10a^2}-\frac{1}{9}},\\
  t(s)=&\;\frac{(s+1)^2(s-2)}{(s-1)^2(s+2)}\,.
  \end{align*}
  Consider the interval $s\in(2,\infty)$ which maps to $t\in(0,1)$. Again only one block contributes
  to the tau function expansion on the corresponding branch at $t=0$. It is explicitly given by
  \ben
  \mathcal{B}\left(2a,a,a,\text{\footnotesize$\frac13$},3a;t\right)=(s_t-1)^{{10a^2}-\frac{7}{72}}
  \left(\frac{s_t+1}{3}\right)^{-{18a^2}+\frac18}\left(\frac{s_t+2}{4}\right)^{{14a^2}-\frac19},
  \ebn
  with $\displaystyle s_t=\frac{(1+\sqrt{t})^{\frac23}+(1-\sqrt{t})^{\frac23}}{(1-t)^{\frac13}}$.
  Note that $t=0$ is not really a branch point of $s_t$.
  \end{eg}
  \begin{eg}
  Taking $\boldsymbol{\theta}=\left(a,a,a,\frac14\right)$, $\boldsymbol{\sigma}=\left(2a,\frac16,\frac16\right)$
  one has a solution (corresponding to Solution~IV in \cite{apvi})
  \begin{align*}
  \tau(t(s))=&\;\mathrm{const}\cdot s^{-{6a^2}+\frac{1}{12}}(s-1)^{-{6a^2}+\frac{1}{12}}(s+1)^{2a^2}(2-s)^{2a^2}
  (2s-1)^{5a^2-\frac{1}{16}},\\
  t(s)=&\;\frac{s^3(2-s)}{2s-1}\,.
  \end{align*}
  Under this parametrization, the interval $s\in(1,2)$ is mapped to $t\in(0,1)$. Tau function asymptotics
  near $t=0$ on the relevant branch
  yields the conformal block
  \ben
  \mathcal{B}\left(a,a,a,\text{\footnotesize$\frac14$},2a;t\right)=
  (s_t-1)^{-{6a^2}+\frac{1}{12}}\left(\frac{s_t}{2}\right)^{-12a^2+\frac{1}{12}}
  \left(\frac{s_t+1}{3}\right)^{{2a^2}}\left(\frac{2s_t-1}{3}\right)^{7a^2-\frac{1}{16}},
  \ebn
  where
  \ben
  s_t=\frac{1}{2}\left(1+u_t+\sqrt{3-u_t^2+\frac{2-4t}{u_t}}\right),\qquad
  u_t=\sqrt{1-\left(4t(1-t)\right)^{\frac{1}{3}}}.
  \ebn
  \end{eg}
  \subsection{Picard solutions}
  The remainder of this section deals with Painlev\'e~VI solutions of Picard type.
  Here the parameters are chosen to be $\boldsymbol{\theta}_{\text{Picard}}=\left(\frac14, \frac14,\frac14,\frac14\right)$. Up to B\"acklund transformations, this is the only case  where the general two-parameter solution of PVI is available \cite{Fuchs,mazzocco}.
  Quite remarkably, for precisely these $\boldsymbol{\theta}$ there exists an explicit formula for $c=1$ conformal block, found by Zamolodchikov \cite{Zamo_AT}. We now briefly explain
  the relation between the two subjects.

   First recall that complex torus can be seen as a two-sheeted covering of the four-punctured
   sphere, its period ratio being determined by the cross-ratio of the punctures. This suggests an
   elliptic parametrization of PVI variable $t$:
  \ben
  q=e^{i\pi\tau}, \qquad
  \tau=\displaystyle\frac{iK'(t)}{K(t)},
  \ebn
  where $K(t)$, $K'(t)$ are complete elliptic integrals of the first kind:
  \ben
  K(t)=\int_0^1\frac{dx}{\sqrt{(1-x^2)(1-tx^2)}},\qquad K'(t)=K(1-t).
  \ebn

  In this parametrization, conformal block with external dimensions corresponding to $\boldsymbol{\theta}_{\text{Picard}}$
  is given by (see Eq. (2.28) in \cite{Zamo_AT})
  \ben
  \mathcal{B}\left(\boldsymbol{\theta}_{\text{Picard}},\sigma;t\right)=
  \frac{\bigl(16t^{-1}q\bigr)^{{\sigma^2}}}{
  (1-t)^{\frac18}\vartheta_3(0|\tau)}.
  \ebn
  Here and below $\vartheta_3(z|\tau)=\sum\limits_{n\in\Zb}e^{i\pi n^2 \tau+2 i n z}$ denotes Jacobi theta function.
  On the other hand, Barnes function duplication identity reduces the formula (\ref{cn}) for the structure constant to
  \ben
  \left(C_n\right)_{\text{Picard}}=\frac{\pi^2 G^4\left(\frac12\right)}{\cos{\pi\sigma_{0t}}}\; (-s)^n 2^{-4(\sigma_{0t}+n)^2},
  \ebn
  The expression for $s$ also simplifies drastically. Jimbo-Fricke relation (\ref{jfr}) implies that for
  fixed $\sigma_{0t}$, $\sigma_{1t}$ the two possible values of $\cos2\pi\sigma_{01}$ are given by
  $-\cos2\pi(\sigma_{0t}\mp\sigma_{1t})$ and therefore $s=-e^{\pm 2\pi i\sigma_{1t}}$.

  Summation over conformal families in (\ref{tauexp}) now gives theta function series  so that we get
  an explicit expression
  \ben
  \tau_{\text{Picard}}(t)=\mathrm{const}\cdot \;\frac{q^{\sigma_{0t}^2}}{t^{\frac18}(1-t)^{\frac18}}
  \frac{\vartheta_3\left(\sigma_{0t}\pi\tau\pm\sigma_{1t}\pi|\tau\right)}{\vartheta_3(0|\tau)}.
  \ebn
  Straightforward check (of the literature \cite{kk}) shows that this function indeed
  satisfies PVI with Picard parameters,
    i.e. the expansion
  (\ref{tauexp}) is complete.
 \vspace{0.2cm}

  The above observations on
  the correspondence between special solutions of Painlev\'e~VI equation and $c=1$
  conformal blocks are collected in the following table:\vspace{0.2cm}
  \begin{center}
  \begin{tabular}{|c|c|}\hline
  Painlev\'e VI solutions & $c=1$ conformal blocks \\ \hline\hline
  Riccati & vertex operators \\ \hline
  Chazy & singular vectors \\ \hline
  Picard & Ashkin-Teller \\ \hline
  algebraic & ? \\ \hline
  \end{tabular}
  \end{center}

 \section{Further questions}
 To summarize, we have obtained a complete series expansion of Painlev\'e VI tau function near
 the singular points $t=0,1,\infty$. This series involves summation over conformal families arising in the OPEs
 of monodromy fields and labeled by $n\in\Zb$, as well as a double sum over Young diagrams which encode
 the contribution of Virasoro descendants.
 The field theory meaning of $\tau(t)$ now becomes clear: \textbf{it is a generating function
 of $c=1$ conformal blocks}.

 This opens a way to intriguing applications of AGT correspondence in the
 theory of monodromy preserving deformations. To increase the number of singular points of
 the linear system (\ref{lsys}), one should deal with AGT representation for the $n$-point conformal block  with $n>4$.
 Increasing rank~$N$ leads to other integer values of central charge, and
 it can be expected that intermediate conformal families will be labeled by $(N-1)$-tuples of integers.
 Another interesting option is the study of isomonodromic tau functions on higher genus Riemann
 surfaces.

 Conversely, isomonodromy problems may provide useful insight
 into CFT. For instance, it would be interesting to understand if conformal blocks
 associated to exceptional algebraic Painlev\'e~VI solutions
 can be computed in explicit form and identify the corresponding theories,
 orbifold CFTs \cite{ginsparg} being the most natural candidates. One may also try to generate
 new explicit examples of conformal blocks for higher values of $c$
 from tau functions associated to branched covers of $\Pb^1$ \cite{korotkin}.

 It is in principle straightforward to obtain from (\ref{tauexp})--(\ref{cn}) similar expansions
 for Painlev\'e~V and Painlev\'e~III tau functions. In particular, this gives full short-distance
 (conformal perturbation theory) expansion of two-point functions in the Ising and
 free-fermion sine-Gordon field theory, and of the PV tau function describing universal part
 of the process of formation of Fisher-Hartwig singularities in the asymptotics of Toeplitz determinants
 \cite{its}.
 This also seems to shed some light on recent results of \cite{abanov1,abanov2}. We hope to return to these issues elsewhere.

\acknowledgments

The authors are grateful to B.~Doyon, P.~Gavrylenko, K.~Kozlowski, A. Marshakov,
  A. Mironov,  V. Roubtsov and V.~Shadura
  for useful discussions and comments. This work  was supported by the  ERC grant no. 279738 -- NEDFOQ (O.~Gamayun),
  the Program of Fundamental Research of the Physics and Astronomy Division of NASU, Joint Ukrainian-Russian SFFR-RFBR project F40.2/108 (N.~Iorgov), IRSES project ``Random and Integrable Models in Mathematical Physics''
  (O.~Lisovyy),  and the joint program of bilateral seminars of CNRS and NASU. A part of this work was done
  when N. Iorgov was a visiting researcher at Tours University.

\end{document}